\documentclass[iop]{emulateapj}
\usepackage{natbib}
\shorttitle{Magnetic Field in AB Aur Disk}
\shortauthors{Li et al.}

\begin{document}

\submitted{Accepted for publication in ApJ}

\title{An Ordered Magnetic Field in the Protoplanetary Disk of AB Aur Revealed by Mid-Infrared Polarimetry}

\author{Dan Li\altaffilmark{1},  Eric Pantin\altaffilmark{1,2}, Charles M. Telesco\altaffilmark{1}, Han Zhang\altaffilmark{1}, Christopher M. Wright\altaffilmark{3}, Peter J. Barnes\altaffilmark{1,4}, Chris Packham\altaffilmark{5,6}, and Naib\'{i} Mari\~{n}as\altaffilmark{1}}
\altaffiltext{1}{Department of Astronomy, University of Florida, 211 Bryant Space Science Center, FL 32611, USA}
\altaffiltext{2}{Service d'Astrophysique CEA Saclay, France}
\altaffiltext{3}{School of Physical, Environmental, and Mathematical Sciences, University of New South Wales, Canberra, ACT 2610, Australia}
\altaffiltext{4}{School of Science and Technology, University of New England, Armidale, NSW 2351, Australia}
\altaffiltext{5}{Physics and Astronomy Department, University of Texas at San Antonio, 1 UTSA Circle, San Antonio, TX 78249, USA}
\altaffiltext{6}{National Astronomical Observatory of Japan, 2-21-1 Osawa, Mitaka, Tokyo 181-8588, Japan}
\email{d.li@ufl.edu}

\begin{abstract}
Magnetic fields (B-fields) play a key role in the formation and evolution of protoplanetary disks, but their properties are poorly understood due to the lack of observational constraints. Using CanariCam at the 10.4-m Gran Telescopio Canarias, we have mapped out the mid-infrared polarization of the protoplanetary disk around the Herbig Ae star AB Aur. We detect $\sim$0.44\% polarization at 10.3 $\micron$ from AB Aur's inner disk ($r<80$ AU), rising to $\sim$1.4\% at larger radii. Our simulations imply that the mid-infrared polarization of the inner disk arises from dichroic emission of elongated particles aligned in a disk B-field. The field is well ordered on a spatial scale commensurate with our resolution ($\sim$50 AU), and we infer a poloidal shape tilted from the rotational axis of the disk. The disk of AB Aur is optically thick at 10.3 $\micron$, so polarimetry at this wavelength is probing the B-field near the disk surface. Our observations therefore confirm that this layer, favored by some theoretical studies for developing magneto-rotational instability and its resultant viscosity, is indeed very likely to be magnetized. At radii beyond $\sim$80 AU, the mid-infrared polarization results primarily from scattering by dust grains with sizes up to $\sim$1 $\micron$, a size indicating both grain growth and, probably, turbulent lofting of the particles from the disk mid-plane.
\end{abstract}

 
\section{Introduction}

Magnetic fields (B-fields) play an important role in star formation. They regulate the gravitational collapse and fragmentation of molecular cores, thus having a strong influence on the global star formation efficiency \citep{dullemond2007,crutcher2012,li2014ppvi}. It can be expected that large-scale B-fields can be dragged inward during core collapse and disk formation, leaving a remnant field in the resultant protoplanetary disk. For a weakly magnetized protoplanetary disk, magneto-hydrodynamic (MHD) turbulence arising from magneto-rotational instability (MRI) is thought to be the primary source of disk viscosity, a crucial driving force for disk evolution and planet formation \citep{balbus1998,turner2014}. Despite this consensus, observations that constrain B-field properties (geometry and strength) in protoplanetary disks are virtually non-existent.

Dichroic emission and absorption of aligned elongated grains produce linear polarization that can trace the B-field morphology. In particular, polarimetric observations of dust thermal emission at centimeter or millimeter wavelengths with single-dish telescopes (e.g., CSO and JCMT) or interferometric arrays (e.g., JVLA, SMA, BIMA, and CARMA) have been used to map B-field structure in young stellar objects (YSOs) at scales from $\sim$50 to thousands of AU (see \citealt{crutcher2012} for a review). However, due to the limited sensitivity and angular resolution offered by current facilities, most of those studies have been focused on B-fields in molecular clumps and cores or Class 0-I objects \citep[e.g.,][]{qiu2013,zhang2014,davidson2014,segura-cox2014,rao2014,liu2016}, rather than classical protoplanetary (i.e., Class II) disks. Using CARMA, \citet{stephens2014} spatially resolved the HL Tau protoplanetary disk in polarized light at 1.3 mm. Their best-fit model was consistent with a highly tilted (by $\sim$50\degr\, from the disk plane) toroidal B-field threading the disk. However, this conclusion is challenged by recent follow-up studies, which show that the 1.3 mm polarization of HL Tau could also arise solely from dust scattering \citep{kataoka2015,yang2016}.

Mid-infrared (mid-IR) polarimetry provides an alternative or complementary approach to the study of B-fields in YSOs and disks \citep{smith2000,barnes2015}. With 8-10-m telescopes, mid-IR observations can achieve 0\farcs3-0\farcs4 angular resolution in the 10-$\micron$ band under most observing conditions, sufficient to map out B-field structure in nearby disks at sub-disk (40-50 AU) scales. Protoplanetary disks are generally thought to be optically thick in the mid-IR out to hundreds of AU from the star \citep{chiang1997}. Hence, mid-IR polarimetry usually probes the emitting particles and B-field near the disk surface (also called the disk atmosphere) rather than its interior. This thin and warm surface layer and layers immediately adjacent to it are a potentially important channel for accretion and angular momentum transfer, since the disk mid-plane at the same radius may be too cold and too well shielded from ionizing radiation to enable MRI (i.e., the ``dead zone'', \citealt{gammie1996}).

To gain new insight into B-fields in protoplanetary disks, we observed AB Aur (HD 31293, MWC 93) with CanariCam \citep{telesco2005,packham2005}, the facility mid-IR camera of the 10.4-m Gran Telescopio Canarias (GTC). AB Aur is an archetypal Herbig Ae star (i.e., intermediate-mass pre-main-sequence stars of 2-4 $M_{\odot}$) at the distance of 144 pc \citep{dewarf2003}. At 4$\pm$1 Myr old, this source still shows evidence of significant accretion ($\sim$10$^{-7}$ $M_{\odot}$ yr$^{-1}$; \citealt{dewarf2003,tang2012}). AB Aur is surrounded by a prominent disk, with mid-IR and 1.3 mm dust emission detected out to $\sim$280 AU and CO line emission detected out to $\sim$500 AU from the star \citep{marinas2006,tang2012}. In both CO and near-IR scattered-light images, the disk is rich in morphological features such as spiral arms and gaps, suggesting a dynamical disk environment and, perhaps, on-going planet formation \citep{pietu2005,hashimoto2011,tang2012}. Previous observations at various wavelengths gave a fairly consistent disk inclination of 27\degr\, (where 0\degr\, corresponds to pole-on), with the major axis of the disk oriented at position angle (P.A.) of 70\degr\, (measured E from N) \citep{pietu2005,tang2012,rodriguez2014}. $H$-band (1.6 $\micron$) polarization of the AB Aur disk has been imaged by \citet{hashimoto2011}, showing a clear centrosymmetric pattern indicative of scattering, as expected at these short IR wavelengths.

The paper is organized as follows. Section \ref{sec:observations} describes our data acquisition and reduction, with results presented in Section \ref{sec:results}. Disk models are presented in Section \ref{sec:analysis}. The implications of our study are discussed in Section \ref{sec:discussion}, with our findings summarized in Section \ref{sec:conclusions}.

\section{Observations and data reduction}\label{sec:observations}
We observed AB Aur on 2015 Feb 6 UT using the 10-$\micron$-band dual-beam polarimetry mode of CanariCam. We integrated on AB Aur for 360 s (on-source) in the $Si$-4 filter ($\lambda=10.3$ $\micron$, $\delta\lambda=0.9$ $\micron$). We chose this filter because it is one of CanariCam's most sensitive filters within the 8-13 $\micron$ atmospheric transmission window, and the spectral energy distribution of AB Aur has a strong silicate emission feature roughly centered at 10 $\micron$, which should provide the best signal-to-noise ratio. For flux and point-spread-function (PSF) calibration, we observed the standard star HD 31398 prior to AB Aur.

In the dual-beam polarimeter mode, a Wollaston prism in the optical path divides incoming light into two beams (ordinary and extraordinary), which are recorded by the detector simultaneously. During integration, a half-wave plate (HWP) in the optical beam rotates to four positions (0\degr, 22\fdg5, 45\degr, and 67\fdg5), which rotates the incoming polarization by 0\degr, 45\degr, 90\degr, and 135\degr. This procedure results in two separate estimates of the fractional Stokes parameters ($q=Q/I$ and $u=U/I$) per full HWP rotation. Using the so-called ratio method to determine $q$ and $u$, small responsivity differences of the detector are cancelled for each HWP setting \citep{tinbergen1996book}. The linear polarization $p$ is calculated as $\sqrt{q^2+u^2-\sigma^2}$ and the P.A. as $0.5*\mathrm{arctan}(u/q)$. Here, $\sigma$ is the noise of $q$ and $u$. This term is introduced to remove the positive bias in $p$ resulting from the noise in the signal.

We computed Stokes $I$, $Q$, and $U$ images from the raw data using $iDealCam$ \citep{li2013idealcam}. The instrumental polarization (IP) was 0.89$\pm$0.05\%, as measured with HD 31398, and was subtracted from the observations of AB Aur in the $Q$-$U$ plane (see Appendix \ref{app:ip} for more details). We note that at the disk center, where the highest sensitivity is achieved, the uncertainty in $p$ is dominated by the uncertainty associated with the IP correction ($\pm$0.05\%). 

The raw data obtained for AB Aur consisted of 80 frames, which permitted us to check for inconsistencies and anomalies within the data. We divided the data into a number of subsets and reduced them separately. Results from different subsets were in good agreement with each other, which ruled out the possibility that the net polarization detected from AB Aur arose as a result of short-term seeing or pointing fluctuations, either of which could result in movement of the PSF of a bright, compact emission source during the exposure.

\section{Results}\label{sec:results}
\begin{figure}
\epsscale{1.2}
\plotone{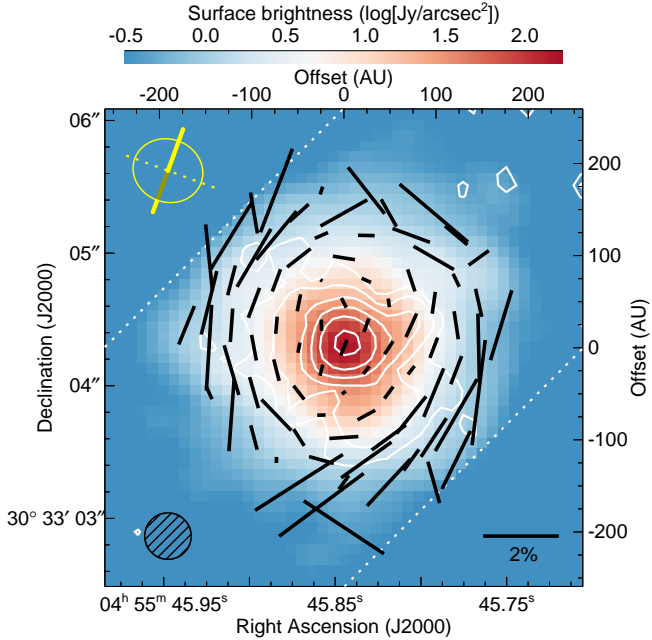}
\caption{The polarization map of the AB Aur protoplanetary disk at 10.3 $\micron$. Displayed in color is the total intensity image of the disk, superimposed by white contours of polarized intensities at 20, 40, 80, 160, 320 and 640 mJy arcsec$^{-2}$. Each polarization vector is derived from an aperture of 3$\times$3 pixels in the original image. Polarization vectors are only plotted where the signal-to-noise (S/N) ratio is higher than 150 in the total intensity image, yielding a maximum uncertainty in the degree of polarization ($p$) of $\sim$1\%. Near the disk center, where the highest S/N ratio is reached, the typical uncertainty in $p$ is $\sim$0.1\%. Angular resolution of the observation is 0\farcs35, as shown in the bottom-left. The upper-left sketch shows the projected spin axis (thick line) and major axis (dashes) of the disk. In the dual-beam polarimeter mode, the effective field of view of CanariCam is a long rectangle $\sim$2\farcs7 in height, as indicated by the two dotted lines. See the electronic edition of the Journal for a color version of this figure.\label{fig:data}}
\end{figure}

\begin{figure}
\epsscale{1.0}
\plotone{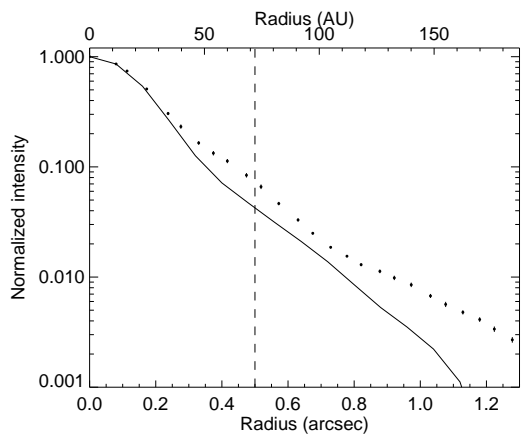}
\caption{Profiles of azimuthally averaged normalized intensity for AB Aur (dots) and the reference PSF star HD 31398. The vertical line is drawn at the radius of 0\farcs5 ($\sim$70 AU), which is the defined boundary between AB Aur's ``inner disk'' and ``outer disk'' as described in the text. \label{fig:profiles}}
\end{figure}

The 10.3-$\micron$ polarization map centered on AB Aur is shown in Fig. \ref{fig:data}. The angular resolution achieved in our observation is 0\farcs35, or 50 AU at 144 pc, corresponding to the full-width at half-maximum intensity of the profile for the PSF standard HD 31398. Extended emission from the disk of AB Aur is spatially resolved (Fig. \ref{fig:profiles}), confirming previous results of \citet{marinas2006}. Linear polarization is clearly detected out to 1\farcs2 (170 AU) from the star. The azimuthally averaged $p$ increases gradually from 0.44$\pm$0.05\% near the star to 1.4$\pm$0.4\% at 170 AU. Polarization vectors ($p$-vectors) within the radius of $\sim$0\farcs5 (70 AU, the ``inner disk'') are oriented almost uniformly with a mean P.A. of 163$\pm$3\degr, a pattern defined by about a half-dozen resolution elements. In contrast, between 0\farcs5 and 1\farcs2 from the star (70 $< r <$ 170 AU, the ``outer disk''), the configuration of $p$-vectors is clearly centrosymmetric.

Although the disk of AB Aur is not precisely pole-on, we do not observe any significant elongation in the total intensity (i.e., Stokes $I$) image, nor do we see any prominent gaps or spiral arms resembling the features observed at other wavelengths \citep{hashimoto2011,tang2012}. However, there is some structure evident in the polarized intensity ($PI$) contours, as shown in Fig. \ref{fig:data}, which are elongated roughly along the major axis of the disk.

\section{Analysis}\label{sec:analysis}
\subsection{Source of polarization}\label{subset:source_of_polarization}
For several widely considered dust alignment mechanisms such as radiation alignment torque (RAT), elongated grains are aligned with short (spin) axes parallel to the B-field lines \citep{lazarian2007,andersson2015}. Consequently, polarization due to absorption of background flux is in the same direction as the B-field, whereas polarization due to emission is orthogonal to it. Hence, the polarization process must be clarified to eliminate this 90\degr\,ambiguity.

Generally, mid-IR polarization along a particular line of sight can arise from dichroic emission, absorption, or both \citep{smith2000,aitken2004}. (Note that the process of scattering is often regarded as negligible in the mid-IR, but we will consider this process below). In the case of AB Aur, however, the contribution of dichroic absorption to the total polarization must be very small, because the disk of AB Aur is close to pole-on, with no evidence of significant mid-IR absorption from intervening material in an envelope or the interstellar medium (ISM) \citep{acke2004,marinas2006}. Supporting this conclusion is that the visual extinction toward AB Aur is only 0.25 mag \citep{roberge2001}. Using the interstellar extinction law from \citet{rieke1985} and the empirical relation between absorptive polarization and the optical depth \citep{smith2000}, we estimate that this level of visual extinction translates into polarization values of $\sim$0.04\% at 10.3 $\micron$, a factor of 10 less than the observed value (0.44\%).

The centrosymmetric component of $p$-vectors in AB Aur's outer disk is remarkably similar to the observation at 1.6 $\micron$ (i.e., Fig. 3 in \citealt{hashimoto2011}), implying that, at least for the outer disk, polarization from scattering probably contributes significantly to the total polarization we observed there. This is an unexpected result, contradicting most, if not all, previous studies on mid-IR polarization in YSOs, where polarization due to scattering is found, or  assumed, to be negligible near 10 $\micron$ \citep[e.g.,][]{aitken1997,smith2000,aitken2002,barnes2015}.

To test our conclusions that the observed polarization of the protoplanetary disk of AB Aur is a mixture of dichroic emission and scattering, we consider radiative transfer models of the disk that take into account both of these polarization mechanisms in the mid-IR.

\subsection{Models}\label{subsec:model}
We modeled the disk of AB Aur using the radiative transfer code RADMC-3D\footnote{http://www.ita.uni-heidelberg.de/\textasciitilde dullemond/software/radmc-3d/}. We assumed a smooth disk with no gaps or envelope. The surface density profile of the disk follows the relation

\begin{equation}\label{eq:surface_density}
\Sigma(r)\propto r^{-q}
\end{equation}
between $r_{\mathrm{in}}$ and $r_{\mathrm{out}}$. The disk inner radius $r_{\mathrm{in}}$, set by the dust sublimation temperature of $\sim$1,500 K, is 0.5 AU \citep{dullemond2001}, and the outer radius $r_{\mathrm{out}}$ is fixed at 400 AU, the lower limit of the disk's radial extension suggested by previous observations \citep{tang2012}. Note that the exact value of $r_{\mathrm{out}}$ has little influence on the results. The disk is flared, i.e., its scale height is described by 

\begin{equation}\label{eq:scale_height}
h(r)=h_{\mathrm{0}}(r/r_{\mathrm{0}})^\gamma,
\end{equation}
with $\gamma>1$. The exponents ($q$ and $\gamma$) in Eqns. \ref{eq:surface_density} and \ref{eq:scale_height}, as well as other key parameters of the model, are collected from the literature \citep{robitaille2007,tannirkulam2008,perrin2009,dullemond2010,tang2012} and summarized in Table \ref{tab:parameters}.

\begin{deluxetable}{ccc}
\tabletypesize{\scriptsize}
\tablecaption{Model parameters\label{tab:parameters}}
\tablewidth{0pt}
\tablehead{
\colhead{Parameter} & \colhead{Value} & \colhead{Unit} 
}
\startdata
$T_*$ & 10,000 & K \\
$R_*$ & 2.5 & $R_{\odot}$ \\
Inclination & 27 & Degree \\
$r_{\rm in}$ & 0.5 & AU \\
$r_{\rm out}$ & 400 & AU \\
$q$ & 1.2 & \nodata \\
$r_{0}$ & 100 & AU \\
$h_{0}$ & 8.5 & AU \\
$\gamma$ & 1.125 & \nodata \\
$M_{\rm dust}$ & 1.2e-4 & $M_{\odot}$ \\
$a_{\rm min}$ & 0.01 & $\micron$ \\
$a_{\rm max}$ & 1.0 & $\micron$ 
\enddata
\tablecomments{Emission of the star is assumed to be blackbody.}
\end{deluxetable}

Dust properties (composition, size distribution, etc.) play a critical role in the models. For computing the dust temperature distribution due to stellar heating and the polarization resulting from scattering, we assumed a homogeneous population of spherical dust particles made of astronomical silicates \citep{draine1984} and spread across the entire disk, with their absorption and scattering mass opacities ($\kappa_{\mathrm{abs}}$ and $\kappa_{\mathrm{sca}}$) calculated with Mie theory. The size distribution of the dust follows the power-law relation

\begin{equation}\label{eq:size_distribution}
n(a)\propto a^{-3.5}
\end{equation}
between $a_{\mathrm{min}}$ and $a_{\mathrm{max}}$, where $a_{\mathrm{min}}$ is set to 0.01 $\micron$, a value appropriate for the ISM grains and used in similar studies \citep{mathis1977,cho2007}. 

The initial value of $a_{\mathrm{max}}$ in the model is obtained using a method developed by \citet{kataoka2015}, as follows. First, we plot the product of $P$ and $\omega$ as a function of $a_{\mathrm{max}}$, where $\omega$ is the dust albedo: $\omega=\kappa_{\mathrm{sca}}/(\kappa_{\mathrm{sca}}+\kappa_{\mathrm{abs}})$. $P$ is the degree of polarization at 90\degr\, due to single scattering: $P=-Z_{\mathrm{12}}/Z_{\mathrm{11}}$, where $Z_{\mathrm{11}}$ and $Z_{\mathrm{12}}$ are elements in the scattering matrix for $\theta_{\mathrm{sca}}=90$\degr. Both $P$ and $\omega$ are averaged over the size distribution. The initial value of $a_{\mathrm{max}}$ was that which made $P\times\omega\approx1.4\%$, i.e., the maximum $p$ we observed in the outer disk of AB Aur. Since the disk of AB Aur is neither flat nor precisely pole-on, the scattering angle is not exactly 90\degr\,everywhere. This results in uncertainty in the phase function. Therefore, during the modeling, we adjusted the value of $a_{\mathrm{max}}$ until the model's level of polarization from the outer disk (70 $< r <$ 170 AU) matched the observed polarization.

\subsection{Adding dichroic emission/absorption}\label{subsec:dichroic}
To implement dichroic emission and absorption into the model, elongated grains are needed. We assumed a single population of oblate spheroids made of astronomical silicates, whose equivalent size (see \citealt{draine1994} for definition) follows the same distribution described by Eq. \ref{eq:size_distribution}. Oblate grains have different cross sections for incident radiation with the electric vector perpendicular and parallel to the grain's symmetry (short) axis. The difference between the two cross sections determines the upper limit on the polarization that can arise from such a grain. The absorption and scattering coefficients ($Q_{\mathrm{abs}}$ and $Q_{\mathrm{sca}}$) for these two orthogonal directions are calculated using the DDSCAT numerical code \citep{draine1994}, with the axis ratio of the oblate grains fixed at 1.5, a value considered to be reasonable by theoretical studies \citep{cho2007}. To compute the amount of polarized emission and absorption produced by a population of such oblate grains embedded in a disk, we use the ray-tracing algorithm presented in \citet{davidson2014}. We assume that the grain's spin (short) axis is parallel to the B-field, as one might expect for RAT. Three-dimensional structure of the B-field is described by three components, $B_x$, $B_y$, and $B_z$, using formulas given in \citet{aitken2002} for a range of field configurations.

We constructed the polarization maps from the model as follows (Zhang et al., in prep.). Initially, RADMC-3D computes the dust temperature and Stokes parameters assuming only spherical grains (as described in Section \ref{subsec:model}). These initial images include only (unpolarized) thermal emission and (polarized) scattered emission. With this temperature distribution in place, the spherical grains are replaced with oblate spheroids. The ray-tracing program is then carried out to compute the polarized emission and absorption along each line-of-sight covering the entire modeling space. Contributions to the polarized light from polarized emission and absorption are added to the $Q$ and $U$ images obtained in the first step. Finally, $I$, $Q$, and $U$ images are smoothed by convolving with a PSF kernel to match the actual spatial resolution (0\farcs35) and then combined to generate $p$-vector maps for comparing with the observations.

Note that in the second step above, all spherical grains are replaced with oblate spheroids. This may not be the case in a real disk, where the dust population is most likely to be a mixture of spherical and non-spherical grains. However, in the computation of polarized emission and absorption, spherical grains are not distinguishable from unaligned oblate grains: the inclusion of spherical grains would have the same effect on the results as a reduced alignment efficiency of oblate grains. In our simulations, the alignment efficiency is described by the Rayleigh reduction factor $R$ (see discussion in Section \ref{sec:best-fit}). Hence, although we do not explicitly include spherical grains in the computation of polarized emission and absorption, their effect has been taken into account in the modeling process.

\subsection{The best-fit model}\label{sec:best-fit}
To search for models compatible with the data, we added B-fields of various configurations into the basic disk model described in Section \ref{subsec:model}, and then examined the resultant polarization maps. We considered all axisymmetric fields discussed in \citet{aitken2002}, including standard poloidal (i.e., all field lines are parallel to each other and also to the spin axis of the disk), standard toroidal (all field lines are circular and parallel to the disk plane), hourglass-shaped, helically twisted, ``K\"{o}nigl,'' and dipole (bipolar) fields. We also considered tilted B-fields whose symmetric axes are not parallel with the spin axis of the disk. The goodness of fit was first evaluated visually by superimposing the model with the data in order to narrow down the parameter space. Then a quantitative comparison was conducted to determine the best-fit model. Degeneracies and limitations in the model are discussed later in this section, with more details in Appendix \ref{app:degeneracy}.

\begin{figure}
\epsscale{1.2}
\plotone{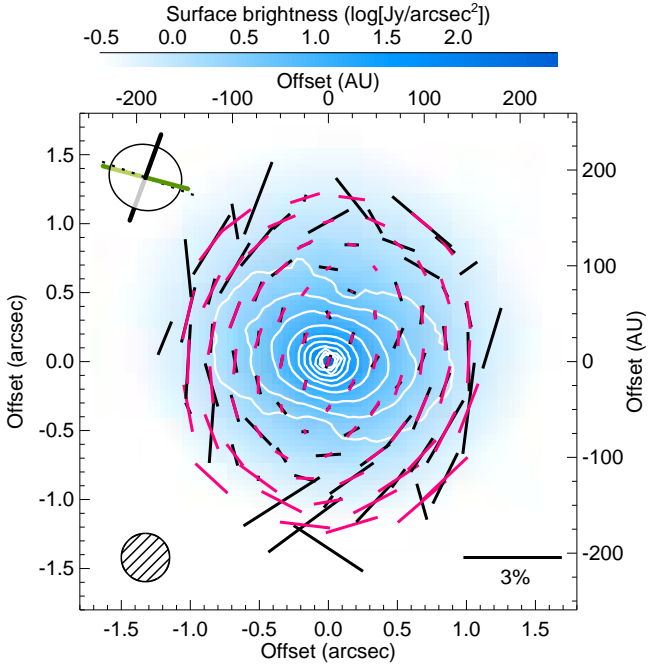}
\caption{The best-fit model (red vectors) superimposed on the observation (black vectors). Displayed in the background is the surface brightness of the model disk, superimposed by model polarized intensity contours. In this model, the disk is threaded by a tilted poloidal B-field, the projected orientation of which is shown in the upper-left sketch (green line). See the electronic edition of the Journal for a color version of this figure.\label{fig:model}}
\end{figure}

\begin{figure}
\epsscale{1.2}
\plotone{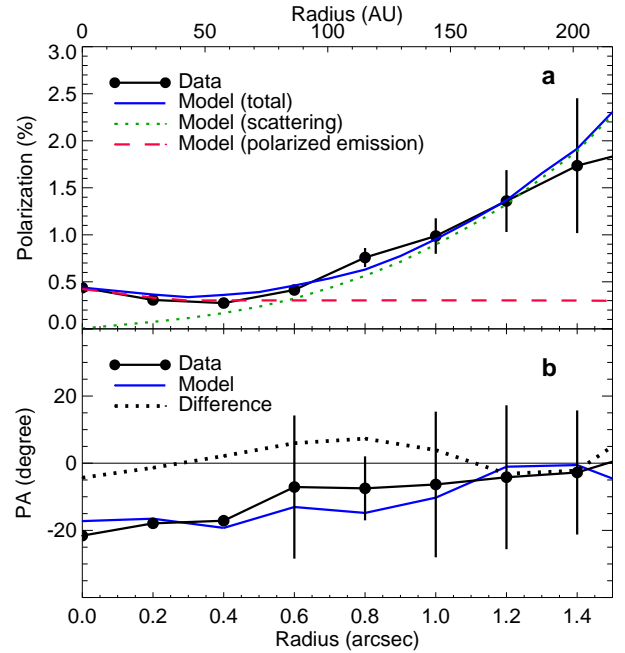}
\caption{Illustration of the goodness of fit. Azimuthally averaged degree of polarization (a) and polarization P.A. (b) of the model (blue lines) and the observation (black lines with 1-$\sigma$ error bars) are compared at a range of deprojected distances from the star. In the outer disk ($r>$ 0\farcs5), most polarization is contributed by scattering (green dotted line). Toward the inner disk ($r<$ 0\farcs5), scattered polarization becomes negligible and polarized emission (red dashed line) from aligned dust grains dominates. See the electronic edition of the Journal for a color version of this figure.\label{fig:fit}}
\end{figure}

The model that fits the observation best is shown in Fig. \ref{fig:model}. It succeeds in reproducing all major features we observed, including the centrosymmetric pattern of $p$-vectors in the outer disk and a more aligned pattern nearer the star. The model strongly supports the conclusion that the mid-IR polarization from the disk of AB Aur contains both dichroic emission and scattered polarization. For the inner disk, thermal emission at 10.3 $\micron$ overwhelms the scattered light by a factor of 100-1000, and consequently the total polarization is dominated by the dichroic thermal emission from aligned dust. At larger radii, thermal emission drops much faster than does the scattered light, and thus, polarization arising from scattering becomes increasingly significant (Fig. \ref{fig:fit}). In the mid-IR, most photons available for scattering are emitted from the unresolved innermost part ($r < 20$ AU) of the disk, including the very hot ($\sim$1,500 K) and compact ($r\approx0.5$ AU) disk inner rim \citep{dullemond2007}. With this illumination geometry, polarization due to scattering would show the signature centrosymmetric pattern, exactly as we observed in the outer disk of AB Aur.

The model supports the assumption that absorptive polarization is negligible. Hence, the mean P.A. of the polarization across the central 0\farcs5 implies that the projected B-field of the inner disk is oriented at P.A. $\approx$ 73\degr\, and (probably coincidently) roughly aligned with the disk's major axis. We found that none of the axisymmetric configurations mentioned above could reproduce this field geometry unless some degree of field tilt was applied. However, adding the field inclination as another free parameter greatly increased the size of the parameter space, which motivated us to only focus on the two simplest configurations: tilted poloidal (straight field lines parallel to each other but not aligned with the spin axis of the disk) and tilted toroidal (circular field lines with planes not parallel to the disk plane). With this constraint, the best-fit model we found is for a poloidal field tilted from the spin axis of the disk by $\sim$33\degr. 

The polarization P.A. indicates the \emph{projected} field orientation, but it does not put direct constraints on the amount of field tilt (i.e., the angle between field lines and the spin axis of the AB Aur disk). However, the radiative transfer simulations allow us to estimate the amount of tilt using not only the polarization P.A. but also the degree of polarization, with some assumptions on the dust shape and the dust alignment efficiency in B-fields. In our models, the former is characterized by the axis ratio of the grains, while the latter is quantified by $R$, the Rayleigh reduction factor \citep{greenberg1968}: grains completely aligned in the field with their spin axes parallel to the field lines have $R=1$, while randomly oriented grains have $R=0$. The field inclination (33\degr) in the best-fit model is obtained by assuming $R=0.05$ and the axis ratio of dust grains to be 1.5. We find that the derived field inclination does not strongly depend on $R$: it only varies moderately between $\sim$30\degr\, to $\sim$40\degr\, when the value of $R$ changes from 0.3 to 0.03. However, because of these additional assumptions and model degeneracies (see Appendix \ref{app:degeneracy}), this value of 33\degr\, is not a precise measurement of the field inclination. Rather, it represents a range ($\sim$30\degr-40\degr) of field inclinations that are consistent with the data. Despite degeneracies and uncertainties in detailed values, we believe the conclusion that the B-field must be tilted is robust.

Starting from the standard poloidal configuration (i.e., field lines oriented perpendicular to the disk plane), the tilted poloidal field in the best-fit model can be obtained as follows. First, the field is tilted by 27\degr\, toward the observer along the P.A. = 160\degr\, (i.e., the projected spin axis of the disk) so that the field lines are now parallel to the line-of-sight (i.e., the observer is now looking along the field lines). Then, the field is tilted again by 20\degr\, along the P.A. $\approx$ 73\degr\, (or 253\degr, which is not distinguishable in this model). At this point, we are viewing a poloidal field almost, but not exactly, pole-on, with the observed polarization resulting from that relatively small component of the B-field projected on to the plane-of-the-sky (Fig. \ref{fig:model}).

Estimating the field inclination in the best-fit model was aided by plotting the azimuthally averaged $p$ and P.A. against the deprojected distance to the star (Fig. \ref{fig:fit}). We did not conduct the conventional $\chi^2$ minimization, because it was too computationally expensive to cover a large range of field inclination. However, the current ``best-fit'' model (reduced $\chi^2=1.71$) should be sufficiently close to the one that minimizes $\chi^2$.

The best-fit model is also compatible with a tilted dipole (or hourglass) field if the scale of the field is large enough that its central part approximates a tilted poloidal field. On the other hand, the tilted toroidal field is probably precluded by the observations, because it cannot reproduce the polarized intensity contours elongated roughly along the major axis of the disk (Fig. \ref{fig:data}). Moreover, a tilted toroidal field would require a high and probably unphysical field inclination ($\ge$50\degr\, from the disk plane) to reproduce the mean polarization P.A. of the inner disk. Our analysis indicates the simplest field geometry that is consistent with the data, but, of course, the actual B-field configuration can be much more complex (e.g., with both poloidal and toroidal components) than the one we present here.

In the ``best-fit'' model described above, the B-field is assumed to thread the entire disk of AB Aur. However, we also considered a highly compact B-field confined to the innermost part ($r<10$ AU) of the disk. This B-field has the same orientation with respect to the disk as the best-fit field, but it only penetrates the central 20 AU-diameter region. Polarized thermal emission arising from such a B-field would be unresolved by CanariCam, but it could appear to extend to radii larger than 10 AU due to PSF smoothing. This model B-field configuration therefore permits us to access in more detail the possibility that the observed central polarized light is not actually resolved spatially. The result from our simulations, which use the observed PSF, implies that, compared to the observed polarization distributions (Figs. \ref{fig:data} and \ref{fig:fit}a), this model predicts polarization that declines too rapidly with radius. For regions immediately outside the central resolution element, this model can only account for $\sim$30\% of the observed degree of polarization. Hence, we conclude that the effect of PSF smoothing is not sufficient to explain the observed polarization from the entire inner disk, and that a highly compact B-field configuration is not favored by our observations.

\subsection{Constraints on dust size}
Our model confirms that the degree of polarization due to scattering depends strongly on the grain size, specifically, the value of $a_{\rm max}$ in Eq. \ref{eq:size_distribution}. For observations near 10 $\micron$, Rayleigh scattering dominates if the grain size is sub-micron. In this regime, the polarization can be as high as 100\% if the scattering angle $\theta_{\rm sca}=90\degr$, but sub-micron-sized grains have very small albedo at 10 $\micron$, and therefore their contribution to the observed polarized light is small. On the other hand, if the grains are too large (e.g., comparable to the observing wavelength), the degree of polarization due to scattering (i.e., $-Z_{\mathrm{12}}/Z_{\mathrm{11}}$) approaches zero. A quantitative analysis of this size dependence is given by \citet{kataoka2015}. In our model we find that $a_{\rm max}$ must be $\sim$1 $\micron$ to correctly reproduce the level of polarization observed in the outer disk of AB Aur. A higher (lower) $a_{\rm max}$ would make polarization too high (low) to be reconciled with our observations. 

\subsection{Models with only scattering or emissive polarization}
Models with only one polarization component, due to either emission or scattering but not both, have also been considered. If the centrosymmetric pattern observed in the outer disk were from emission by aligned dust grains, the B-field in that region would have a significant radial component, which is in stark contrast to the B-field morphology of the inner disk. We have attempted to reproduce this field geometry with an hourglass-shaped field viewed roughly along its symmetry axis. This results in an approximately poloidal field at the disk center with an increasing radial component toward larger radii. However, it is very difficult for this configuration to match the observations, because the centrosymmetric pattern produced by such a field rapidly disappears when the line-of-sight deviates even slightly from the symmetry axis of the field. In contrast, the same pattern produced by scattered polarization is very robust and able to maintain the symmetric shape when the line of sight is inclined from the spin axis of the disk.

We find that scattering alone is also unable to reproduce all the observations. The disk of AB Aur is close to pole-on, and our model suggests that more than 90\% of the 10-$\micron$ photons available for scattering originate in a small region close to the star ($r<20$ AU) and unresolved at our instrumental resolution (50 AU). Under this illumination geometry, scattered emission would produce a clear centrosymmetric polarization pattern, not only in the outer disk but also in the inner disk, which is not observed. Fig. \ref{fig:fit} shows that the degree of polarization is more or less ``flat'' inside about 70 AU from the star.  This is hard to explain with scattered emission that, according to our model, would result in decreasing polarization toward the disk center.

In the near-IR, a ``polarization disk'' may result entirely from scattering in the envelope of a YSO \citep{whitney1992,whitney1993}. However, in this scenario, multiple scattering (mainly double scattering) produces polarization vectors along the disk's major axis, which is the opposite of what we observed. In addition, our simulations confirm that the intensity of single scattering is already very small compared to thermal emission in the inner disk of AB Aur, and multiple scattering is essentially negligible.

It is shown by \citet{yang2016} that a uniform polarization pattern aligned with the minor axis of a disk can be produced by self-scattering, given that the disk is moderately inclined (e.g., 45\degr) and optically thin. The latter is generally true for cm-mm observations of disks. However, the disk interior of AB Aur is optically thick near 10 $\micron$ and the inclination is low (27\degr), so our observations cannot be explained by that mechanism.

\subsection{Polarization from a jet?}
The mean P.A. of the 10.3-$\micron$ polarization across the inner disk of AB Aur is 163$\pm$3\degr. This is very close to the P.A. of a jet known to be associated with AB Aur \citep{rodriguez2014}. Although dust grains may be aligned mechanically in a jet/outflow, we have concluded that the polarization most likely originates on the disk surface rather than at the jet for the following reasons. First, deep mid-IR images suggest that the jet is not detected at 10.3 $\micron$. This indicates that the jet does not significantly contribute to the mid-IR flux  \citep{marinas2006}. Therefore, even if dust grains are aligned in the jet outflow, they are not likely to produce any detectable mid-IR polarized emission. Second, because the disk is optically thick in the mid-IR, if there were any polarization footprints left by the jet, we would only see them on the front side of the disk. This would result in a noticeable asymmetry in the polarization map between the NW and SE parts of the disk, which is not observed.

\section{Discussion}\label{sec:discussion}
The 10.3-$\micron$ polarimetry probes the physically and optically thin, warm surface layer, which contains much less than 0.1\% of the surface density of a protoplanetary disk \citep{takeuchi2003}. However, while thin, this layer is nevertheless of essential importance for MRI to operate in the disk. Using mid-IR polarimetry as a surface-specific probe, our observations reveal the distinctive footprint of an ordered B-field existing near the disk surface, supporting the idea that MRI-induced accretion and angular momentum transport can operate through, or near, this layer.

In the ideal MHD limit, both analytical and numerical studies of non-turbulent cores show that, if the B-field and the spin axes of the core are aligned, the formation of rotationally supported disks is suppressed \citep{galli2006,mellon2008}, at odds with the observed abundance of YSOs surrounded by protoplanetary disks. A large-scale poloidal B-field tilted relative to the protostellar/disk spin axis, along with turbulence and other non-ideal MHD effects (i.e., ambipolar diffusion, the Hall effect, and Ohmic dissipation), has been proposed to alleviate this ``magnetic braking catastrophe'' \citep{li2014ppvi,hennebelle2009,joos2012}. Our observations imply that the misalignment hypothesis may indeed be key to understanding the formation of disks like AB Aur's. A tilted poloidal configuration also means that there is a vertical (i.e., perpendicular to the disk plane) component in the B-field. In numerical studies of MRI-driven turbulence in protoplanetary disks, it is found that, without this vertical component, it is impossible to generate sufficient MRI-driven turbulence and resultant accretion rates that are high enough to be consistent with observations \citep{simon2013,simon2015}.

We note that in RAT \citep{lazarian2007}, a favored mechanism to mutually align dust grain spin axes, there is a critical field strength below which the process cannot work regardless of properties of the radiation field or the grains. On the other hand, MRI can be suppressed if the B-fields are too strong \citep{wardle2007,fromang2013}. If one accepts that both RAT and MRI are able to operate near the disk surface of AB Aur, then some constraints can be imposed on the field strength ($B$), which is otherwise not measurable from our present data. 

Following \citet{hughes2009}, the critical field strength ($B_{\mathrm{min}}$), below which grains cannot be aligned in the scheme of RAT, is (using cgs units)

\begin{equation}
B_{\mathrm{min}}=4.1\times10^{-11}\frac{anT_dT_g^{1/2}}{s^2},
\end{equation}
where $a$ is the grain size, $n$ is the gas density, $T_d$ ($T_g$) is the dust (gas) temperature, and $s$ is the axis ratio of the grain. Considering a grain size of 0.1 $\micron$ and $s$ of 1.5, with other quantities sampled in the model at the layer of unity optical depth at 10.3 $\micron$ ($\tau_{10.3}=1$) and $r=25$ AU (i.e., the half width of the resolution element), where $n\approx8\times10^8$ cm$^{-3}$ (assuming a gas-to-dust ratio of 100 and a mean molecular weight of 7/3) and $T_d=T_g=300$ K, the critical field strength $B_{\mathrm{min}}$ is about 800 $\mu$G (or $8\times10^{-8}$ T).

To enable MRI, the B-field energy density should be lower than the thermal energy density. This criterion can be written as a condition on the plasma parameter $\beta$, the ratio between thermal and magnetic pressure (assuming Keplerian disk) \citep{wardle2007,fromang2013}

\begin{equation}
\beta\equiv\frac{P_g}{P_B}\ge\frac{8\pi^2}{3},
\end{equation}
where $P_B=B^2_{max}/8\pi$, and $P_g=\rho c_0^2=\rho kT_g/\mu m_p$ ($\rho$ the mass density, $c_0$ the isothermal sound speed, $k$ the Boltzmann constant, $\mu$ the mean molecular weight, and $m_p$ the proton mass, all in cgs units). Again, considering $r=25$ AU on the $\tau_{10.3}=1$ surface, we have $B_{max}\approx6$ mG ($6\times10^{-7}$ T). The field strength corresponding to $\beta=1$ (i.e., equipartition between thermal and magnetic energy) should be considered an absolute maximum, which is about 30 mG ($3\times10^{-6}$ T) at the same location.

Therefore, we estimate that, to activate both the RAT and MRI mechanisms, the B-field strength must be of order of 1-10 mG at $r=25$ AU. This field strength is significantly higher than that found for the interstellar B-field ($\sim$10 $\mu$G), but comparable with those measured for much younger protostellar cores ($\sim$10 mG) \citep{crutcher2012}.

Finally, we comment on the scattered polarization detected in the outer disk of AB Aur. While commonly observed in the optical and near-IR regions, polarized scattered light has not been observed previously in the mid-IR for any protoplanetary disks. This polarization component carries no information about the B-field, but it shows that the maximum size is nicely constrained. In particular, our modeling implies that dust grains as large as $\sim$1 $\micron$ are needed on the surface of AB Aur's outer disk to be reconciled with the observation. This particle size is larger than that of typical interstellar grains (0.01-0.1 $\micron$) and likely results from grain growth, a crucial and expected step in the earliest stages of planet formation \citep{laibe2008}. Moreover, since micron-sized or larger particles should settle toward the disk's mid-plane on timescales of $\sim$10$^5$ years \citep{laibe2014}, the presence of such grains near the surface of AB Aur's 4 Myr old disk implies vertical mixing and therefore that significant turbulence may be occurring even in the outer part of a disk like AB Aur's.

\section{Conclusions}\label{sec:conclusions}
We present GTC/CanariCam mid-IR (10.3 $\micron$) polarimetric observations with $\sim$0\farcs5 (50 AU) angular resolution of the protoplanetary disk of AB Aur to explore its magnetic field. The key findings are summarized below:
\begin{enumerate}
\item Linear polarization is detected from the disk of AB Aur out to $r\approx$ 1\farcs2 (170 AU). The polarization map shows two distinct regions, which we call the inner disk and the outer disk. Polarization vectors in the inner disk ($r<70$ AU) are approximately parallel to each other, whereas those in the outer disk (70 $< r <$ 170 AU) form a clearly centrosymmetric pattern. The (azimuthally averaged) degree of polarization increases from 0.44$\pm$0.05\% near the star to 1.4$\pm$0.4\% at 170 AU.
\item We modeled the observations using RADMC-3D with customized code to include polarization from emission and absorption by aligned elongated dust grains. Our results show that the observed polarization is well reproduced when both polarized emission and polarization from scattering are included in the model. In the best-fit model, the disk of AB Aur is threaded by a poloidal field tilted from the spin axis of the disk by $\sim$30\degr-40\degr. Polarization of the inner disk is dominated by dichroic emission from elongated grains aligned in the B-field. In contrast, polarization of the outer disk is largely due to scattering.
\item The disk of AB Aur is almost certainly optically thick at 10.3 $\micron$, so mid-IR polarimetry probes emitting dust grains and the B-field in the warm disk surface. Our observations imply that this surface layer is indeed magnetized, a crucial condition for MRI to operate. Furthermore, we estimate that, for both RAT and MRI to operate, the field strength on the disk surface should be of order 1-10 mG.
\item A poloidal B-field tilted relative to the disk spin axis supports theories requiring such a misalignment to mitigate the ``magnetic breaking catastrophe'' \citep[e.g.,][]{hennebelle2009,joos2012}. It also ensures a considerable vertical component of the field (i.e., perpendicular to the disk plane), which is needed to create a sufficiently high accretion rate through MRI-driven turbulence as suggested by observations \citep{simon2013,simon2015}.
\item Significant polarization arising from scattering in the outer disk of AB Aur requires micron-sized grains near the disk surface, indicating grain growth and possible lofting of these particles to the disk surface by turbulence.
\end{enumerate}

Our study of AB Aur is the first to probe B-fields in a protoplanetary disk with mid-IR polarimetry, and it demonstrates the potential of this technique. While our observations provide critical boundary conditions that must be satisfied by the B-field interior to the disk of AB Aur, that interior B-field geometry remains otherwise undefined. Other observing facilities, such as the Atacama Large Millimeter Array (ALMA) and Very Large Array (VLA) observing at sub-millimeter and centimeter wavelengths, will permit probing of disk interiors all the way to the disk mid-plane, observations that will strongly complement those in the mid-IR.

\acknowledgments
We are grateful to the GTC staff for their support during the queue observations. E.P. acknowledges the support from the AAS through the Chr\'{e}tien International Research Grant and the FP7 COFUND program-CEA through an enhanced-Eurotalent grant, and the University of Florida for its hosting through a research scholarship. C.M.T. acknowledges support from NSF grants AST-0908624, AST-0903672, and AST-1515331. C.M.W. acknowledges financial support from Australian Research Council Future Fellowship FT100100495. This research is based on observations using CanariCam at the Gran Telescopio Canarias (GTC), a partnership of Spain, Mexico, and the University of Florida, and located at the Spanish Observatorio del Roque de los Muchachos of the Instituto de Astrof\'{i}sica de Canarias, on the island of La Palma.

{\it Facilities:} \facility{GTC (CanariCam)}.

\bibliographystyle{apj}

\appendix

\section{A. Instrumental polarization}\label{app:ip}
During the observations presented above, CanariCam was mounted at the Nasmyth-A focal station of the GTC. Because the design of CanariCam ensures a very low level of instrumental polarization (IP), any IP present in the data arises predominantly from the 45\degr\, reflection off the telescope's tertiary mirror. Therefore, the orientation of the IP is a function of telescope pointing and the position angle of the field-of-view (FOV), and is thus a known value (accurate within a few degrees) for every observation. The magnitude of the IP is a function of wavelength, with little temporal variation. Numerous data from the commissioning and scientific operations of CanariCam have confirmed that the IP is very stable. To double-check the IP in our observations, the mid-IR photometric standard HD 31398, a bright giant of spectral type K3 with no expected intrinsic polarization, was observed along with AB Aur.

\begin{figure}
\epsscale{1.0}
\plotone{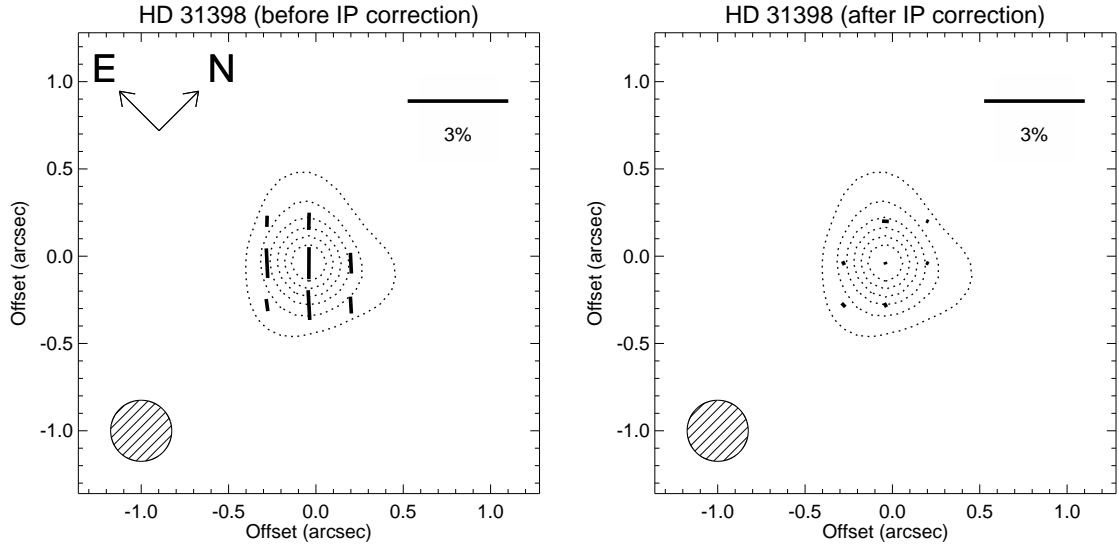}
\caption{The polarization map of the PSF standard HD 31398 at 10.3 $\micron$. Contours are drawn where S/N ratios (in the total intensity image) reach 300, 600, 900, etc. For a dual-beam polarimeter such as CanariCam, these S/N ratios yield absolute uncertainties in the degree of polarization of 0.5\%, 0.25\%, 0.17\%, etc., respectively. Each polarization vector is derived from an aperture of 3$\times$3 pixels in the original image. Angular resolution of the observation is 0\farcs35, as shown in the bottom-left corners. \textit{Left panel}: the polarization map before the IP correction is applied. \textit{Right panel}: the same image after the IP correction.\label{fig:hd31398}}
\end{figure}

The polarimetric image of HD 31398 (the left panel of Fig. \ref{fig:hd31398}) confirmed that both the magnitude (0.89$\pm$0.05\%) and the orientation ($\sim$45\degr) of the IP were in good agreement with the expected values. After IP subtraction using the standard procedure, the polarization of HD 31398 was essentially zero (the right panel of Fig. \ref{fig:hd31398}). The same procedures to correct for IP were then applied to the AB Aur data.

To further verify the accuracy of the IP correction and to look for any residual instrumental effects, we observed AB Aur again on a different night. This additional data set and that presented in Section \ref{sec:results} are consistent with each other within the measurement uncertainties even though they were obtained with different telescope pointing and different FOV position angles. This indicates that the IP correction has been applied properly, and that the polarization we present in Fig. \ref{fig:data} is not instrumental. However, we chose not to combine the two data sets, because poor seeing degraded the angular resolution of the data obtained on the second night.

\section{B. Degeneracies in the model}\label{app:degeneracy}
In addition to the dust properties, scattered light polarization also depends on the disk geometry, i.e., $h_0$ and $\gamma$ in Eq. \ref{eq:scale_height}. These parameters, together with $a_{\mathrm{max}}$, are degenerate. The degree of scattered polarization is a function of $\theta_{\mathrm{sca}}$, the scattering angle. The curve of $\theta_{\mathrm{sca}}$ peaks at 90\degr\, and falls rapidly for $\theta_{\mathrm{sca}}>90$\degr\, and $\theta_{\mathrm{sca}}<90$\degr\, \citep{kruegel2003book,kataoka2015}. For low-inclination disks like AB Aur, a small change in the disk scale height or flaring index corresponds to $\theta_{\mathrm{sca}}$ varying around 90\degr, thus affecting the degree of polarization of scattered light considerably. To simplify the modeling, in the current setup, both $h_0$ and $\gamma$ are treated as fixed parameters, and their values are collected from the literature. We note that allowing $h_0$ and $\gamma$ to change does not change our conclusion that grains significantly larger than typical ISM grains are required, although the exact value of $a_{\mathrm{max}}$ may vary moderately.

In the computation of emissive polarization, there are two previously noted parameters resulting in model degeneracies: the axis ratio of dust particles and the Rayleigh reduction factor $R$. A high degree of polarization can be a result of highly elongated dust grains with low alignment efficiency, or well-aligned particles of axis ratio close to unity. In the scheme of RAT, the efficiency of dust alignment is determined by the dust size, radiation field, and gas density \citep{cho2007}. If $R$ can be estimated from these parameters, it would help break the degeneracy between the dust shape and the value of $R$. However, we did not include this treatment. Instead, we fixed the axis ratio of dust grains to be 1.5 while allowing $R$ to vary in the model. Initially, we attempted to fit the data with a uniform $R$ across the entire disk. When $R=0.05$, we were able to reproduce the polarization map for most parts of the disk reasonably well, but polarization from the innermost pixels was lower than observed. To correct for this, in the best-fit model (Fig. \ref{fig:model}), we relaxed the assumption of a uniform $R$ and allowed for enhanced dust alignment efficiency ($R=0.15$) for the innermost resolution element.

\end{document}